# CALDER: Cryogenic light detectors for rare events search

**L. Pagnanini***
*Gran Sasso Science Institute (INFN), L'Aquila, Italy*
*E-mail:* lorenzo.pagnanini@gssi.infn.it

**I. Colantoni, A. Coppolecchia**
*Università La Sapienza, Rome, Italy*

**F. Bellini, L. Cardani, N. Casali, C. Cosmelli, A. Cruciani**
*Università La Sapienza and INFN, Rome, Italy*

**A. D'Addabbo**
*Laboratori Nazionali del Gran Sasso (INFN), Assergi (AQ), Italy*

**M. G. Castellano**
*Istituto di Fotonica e Nanotecnologie (CNR), Rome, Italy*

**S. Di Domizio**
*Università di Genova and INFN, Genova, Italy*

**C. Tomei, M. Vignati**
*INFN Sezione di Roma, Rome, Italy*

The CALDER project aims at developing cryogenic light detectors with high sensitivity to UV and visible light, to be used for particle tagging in massive bolometers. Indeed the sensitivity of CUORE can be increased by a factor of 3, thanks to the reduction of the $\alpha$-background, obtained by detecting the Cherenkov light (100 eV) emitted by $\beta/\gamma$ events. Currently used light detectors have not the features required to address this task, so we decided to develop a new light detector using Kinetic Inductance Detector as a sensor. This approach is very challenging and requires an intensive R&D to be satisfied. The first results of this activity are shown in the following.

*Speaker.

## 1. Introduction

The future of the bolometric experiments dedicated to neutrinoless double beta decay, is closely linked to the ability of reducing background, using an active tool for particle tagging. The main goal of the CUPID (CUORE Upgrade with Particle Identification) project [1] is to understand what is the best choice for the CUORE [2] upgrade. One of the possibilities is to detect the small amount of Cherenkov light produced in $TeO_2$ crystals by $\beta/\gamma$ events to suppress the flat background from smeared $\alpha$ particles [3]. The figure 1 shows the expected sensitivity of CUORE and the results achievable with Cherenkov tagging in the next generation experiments. The target of the CALDER project [4] is to develop new light detectors with a baseline resolution of 20 eV RMS, that allow to reach a satisfactory background suppression.

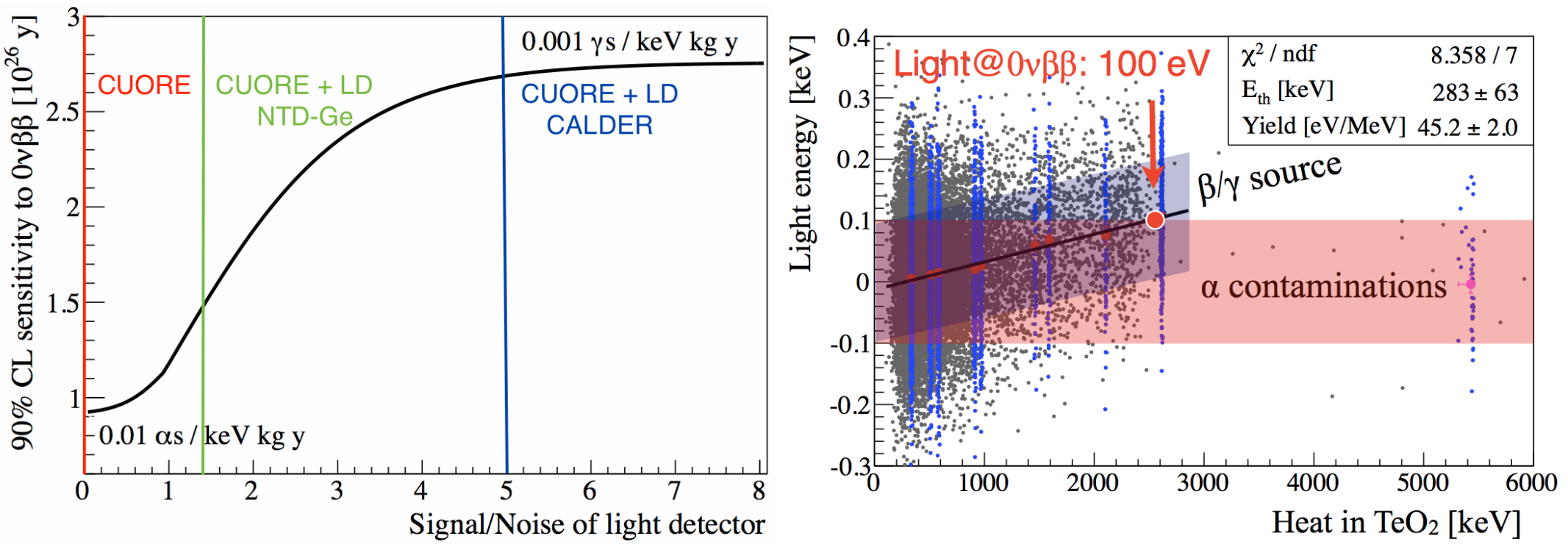

**Figure 1:** *Left.* CUORE sensitivity to the $^{130}$Te $0\nu\beta\beta$ decay half-life as a function of signal to noise ratio of the light detector. The three marked configuration are the CUORE current one (no light detectors), the CUORE with Neutron Transmutation Doped Germanium light detectors one and finally the hypothetical configuration given by application of CALDER light detectors.
*Right.* Detected Cherenkov light versus the heat signal in the $TeO_2$ bolometer for all the acquired events (gray dots) and for the events belonging to the $\gamma$-peaks (blue dots); the light collected for the events is clearly energy dependent (red dots below 3 MeV) and compatible with zero for the $\alpha$-decay of the $^{210}$Po (pink dot at 5.4 MeV); the separation among the $\beta/\gamma$ band (in blue) and the $\alpha$band (in red) is not effective using the Germanium light detector developed in LUCIFER [5]; figures adapted from [3].

## 2. KID sensors for cryogenic light detectors

The detection of the Cherenkov radiation from the CUORE crystals requires to use light sensors with the following features [4]:

- an active area of 5×5 cm$^2$ (area of the surfaces of the CUORE crystals);
- resolution below 20 eV RMS;
- operating temperature 10 mK;
- reproducibility (1000 detectors).

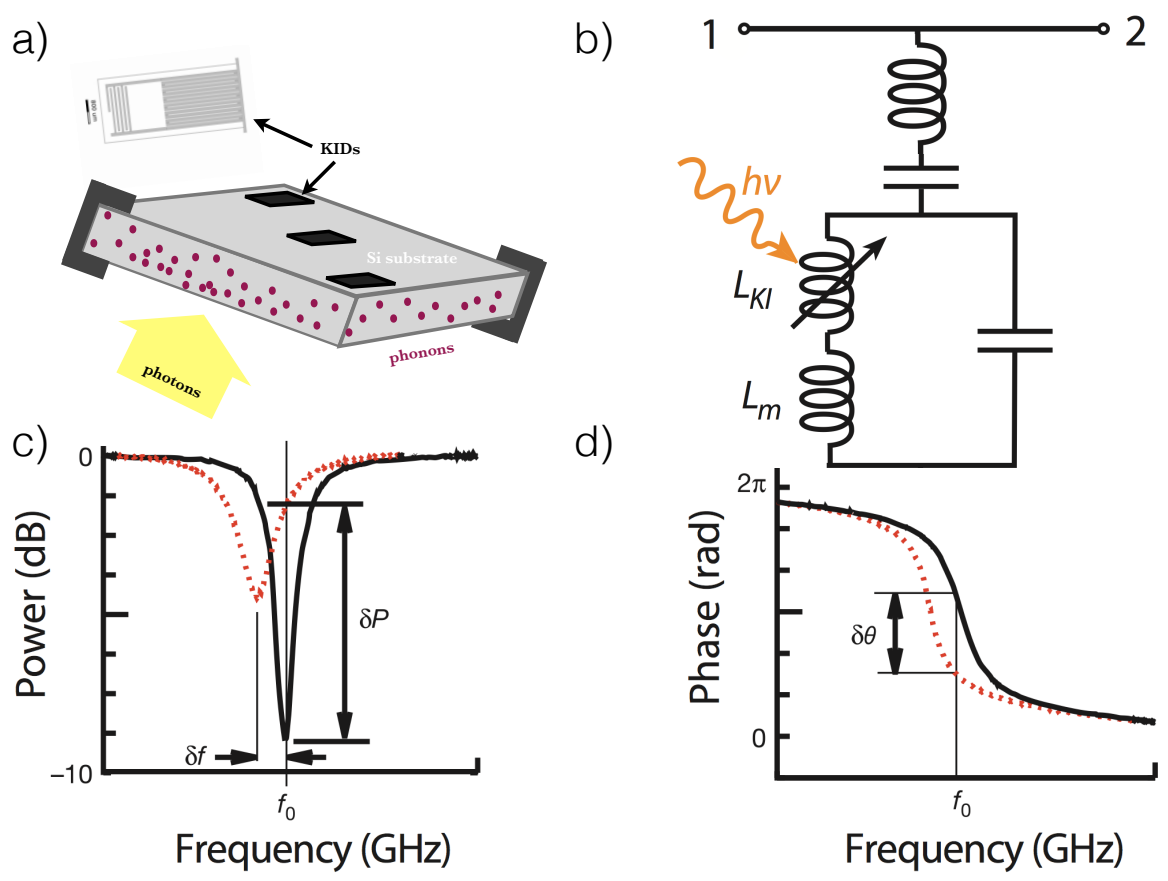

**Figure 2:** *Left.* KID working principle: a) A photon is absorbed by the substrate creating a cascade of athermal phonons. Phonons with sufficient energy (Energy required = fraction of meV) can be absorbed by the superconducting film (MKID), breaking Cooper pairs, creating quasiparticles and varying the kinetic inductance. b) The kinetic inductance $L_{KI}$ acts as a variable inductor in an LC circuit weakly coupled to a microwave feedline. Incident energy that breaks Cooper pairs causes an increase in the kinetic inductance and dissipation in the circuit. c) The increased inductance shifts the resonance to lower frequency, while the increased dissipation decreases the quality factor. d) A phase shift is seen when monitoring the transmission at a given frequency. Figure adapted from [6].

The common light detectors used until now have not all the required characteristics, so we decided to develop a new one adapting the phonon mediated kinetic inductance detector [8]. The working principle of this detectors is well detailed in [4] and summarised in the figure 2.

## 3. First results and future perspectives

Since the beginning of 2014, when the project CALDER started, many prototypes are tested; the single-pixel design [7, 9] has been kept the same one, while some other features (coupling distance, ground plane etc.) have been changed to understand how each parameter plays a role in the global sensitivity definition. The preliminary results from detector consisting in 4 Al pixels deposited on a 2×2 cm$^2$ silicon substate, demonstrated a baseline energy resolution of 230 eV can be achieved (see figure 3); more recent and detailed results will be published in [10]. Further improvements will be possible by optimizing the geometry of the detector and using a more sensitive superconductor, as clearly pointed out by comparing the table 1 with the following formula,

$$\Delta E \propto \frac{T_c}{\sqrt{\tau_{qp} L_{kin}}}, \tag{3.1}$$

where $\Delta E$, $T_c$ and $L_{kin}$ are respectively the energy resolution, the critical temperature of the used superconductor and its kinetic inductance.

Over the 2015 the production of Aluminum sensors at high Q will be completed, reaching 50-100 eV baseline noise, moreover the development of TiN - Ti/TiN sensors will start, reducing the number of pixels per detector and reaching a 20 eV baseline noise. The integration with the

| | $T_c$ [K] | $L_{kin}$ [pH/sq] | $\tau_{qp}$[$\mu$s] |
|---|---|---|---|
| Aluminium | 1.2 | 0.05 | 100-1000 |
| TiN | 0.9 | 3 | 10-200 |
| Ti+TiN | >0.4 | 30 | 50 |

**Table 1:** Critical temperature $T_c$, kinetic inductance $L_{kin}$ and recombination time of the quasiparticles $\tau_{qp}$ for a superconductors used in CALDER.

CUORE/LUCIFER setup at LNGS will begin over the 2017, building a demonstrator with an array of $TeO_2$ bolometers monitored by the new light detectors.

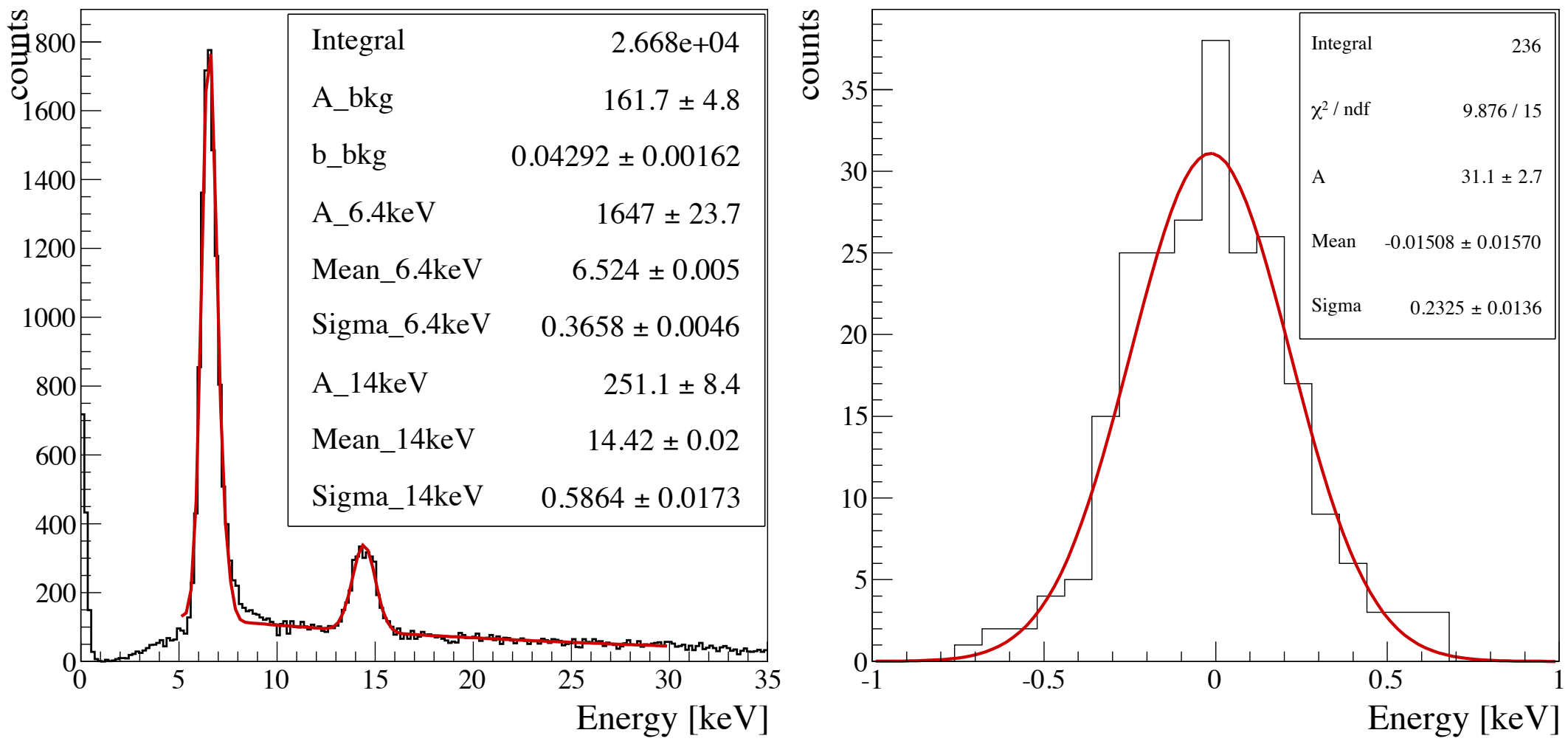

**Figure 3:** *Left.* Sum of the energy released in the four resonators of the tested KID, the fit of the peaks produced by the $^{57}$Co calibration source is also shown. *Right.* RMS energy resolution of the baseline obtained by combining the four pixels.

## Acknowledgements

This work was partially supported by the CALDER experiment, funded by ERC under the European Union's Seventh Framework Programme (FP7/2007-2013)/ERC grant agreement No. 335359.